\documentclass[preprint,aps]{article}

\begin{document}
\begin{titlepage}
\centerline{\large \bf Random phase approximation with exchange for the photoionization}
\centerline{\large \bf of confined atoms: Xe in C$_{60}$ fullerene} 
\vglue .1truein
\centerline{Zhifan Chen and Alfred Z. Msezane}
\centerline{\it Center for Theoretical Studies of Physical Systems,  and
Department of Physics}
\centerline{\it Clark Atlanta University, Atlanta, Georgia 30314, U. S. A.}
\centerline{}
\centerline{}
\centerline{\bf ABSTRACT}
Photoionization of a Xe atom confined inside C$_{60}$ has been studied using the random phase 
approximation with exchange (RPAE) method. The C$_{60}$ fullerene has been described by an 
attractive short range spherical well with potential $V(r)$,  given by $V(r)=-V_0$
for $r_i<r<r_0$, otherwise $V(r)=0$ where $r_i$ and $r_0$ are respectively, the 
inner and outer radii of the spherical shell. The radial parts of the wave function in 
the three regions $r<r_i$, $r_i<r<r_0$ and $r>r_0$ have been obtained by
solving the Schr\"{o}dinger equation using both regular and irregular 
solutions and the continuous boundary conditions at $r_i$
and $r_0$. The photoionization cross sections for the Xe $4d$, $5s$ and $5p$ electrons in  
the Xe@C$_{60}$ endohedral molecule have been evaluated and
compared with those of the photoionization for the free Xe atom and other previous
calculations for the Xe@C$_{60}$ fullerene.  
Our method surmounted the weaknesses of the previous model potential calculations and 
demonstrated significantly stronger correlated confinement resonances for 
the Xe@C$_{60}$ photoionization.  

\centerline{} 

\end{titlepage}
\section{Introduction}
In recent years the photoionization of an atom confined by a C$_{60}$ fullerene 
has received both extensive and intensive research [1-21], because of its importance to  
the development of  
nanotechnology. Endohedral C$_{60}$ molecules can be prepared by accelerating  
ions of atoms and implanting them into the C$_{60}$ cage [22] and by a surgical method [23]. 
In the former method the
ions should have just enough energy to open up the cage and enter.
The endohedral C$_{60}$ molecules can also be achieved by co-evaporation of 
the carbon and the metal in an arc discharge chamber [24]. If the dopant remains
neutral, it normally stays in the centre of the cage, which will greatly simplify the
theoretical treatment of the photoionization process. 
In the past several years there have been a number of theoretical studies of the photoionization of 
endohedral atoms [1-6,11-21], and only very few experimental studies [10] have been reported.
The confining effect 
of the C$_{60}$ cage in the theoretical studies  
is usually modeled by a potential well,  
such as a $\delta$-like potential [11-17] or a spherical, short
range attractive well [18-21]. It should be noted, however that 
in realty the whole space has been divided by the C$_{60}$ potential into several regions. 
In these regions the solution of the Schr\"{o}dinger equation
for a confined atom is usually different from that of a free atom.  
If only the boundary condition is changed  
in solving the Schr\"{o}dinger equation for the confined atom,   
the solution is kept the same as that of the free atom, the continuum wave function will not 
have the phase 
shift due to the photoelectron scattered by the C$_{60}$ potential well [12].   
Because of this care must to be taken to obtain the appropriate wave functions by 
solving the Schr\"{o}dinger equation.

The $\delta$-like potential, which is not zero only in an infinitely thin 
spherical layer, is not a good approximation to
the C$_{60}$ shell. A potential well, which models the confining effect of the 
C$_{60}$ cage has been optimized through the study of C$_{60}$ photoionization [25] and
employed in studies of photoionization [20-21]. 
This potential well 
has an inner radius of 5.75 a.u. which is approximately equal to the radius of the 
C$_{60}$ molecule
and the thickness of the well is approximately 1.89 a.u. The depth of the well 
is -0.3028 a.u. [9, 14].

In this paper we have performed a random phase approximation with exchange (RPAE) calculation 
for the photoionization of the Xe $4d$, $5s$ and $5p$ electrons in the Xe@C$_{60}$ endohedral
molecule. The results have been compared with those of previous calculations 
for the Xe@C$_{60}$ and with both the theoretical and experimental data  
for the free Xe atom.

\section{Theory}
We study the photoionization process of a Xe atom
located in the center of the C$_{60}$ fullerene shell, Xe@C$_{60}$.
The wave function of an optical electron of the Xe atom in this spherically symmetrical 
well has the standard form $\psi(r)=[P(r)/r]Y_{lm}(\theta,\phi)$. 
As $r_i$ is much larger than the electron shell of the Xe atom,
the wave function of an optical electron in the ground state can be considered
approximately to coincide with the corresponding wave function of a free Xe atom [14].
However, the continuum wave function will be quite different from that of the free atom.
The continuum wave function, $P(r)$ satisfies the radial Schr\"{o}dinger 
equation,
\begin{equation}
{1 \over 2}[P(r)''-{l(l+1) \over r^2} P(r)]+[{k^2 \over 2}-V(r)-U(r)]P(r)=0,
\end{equation}
where $k$ is the momentum of the photoelectron, $k^2=2E$, E is the energy of the photoelectron
and $l$ is the orbital angular momentum.  $U(r)$ is the self-consistent field 
created by the
atomic nucleus and all the atomic electrons, acting upon the
optical electron and
$V(r)$ is the model potential of the C$_{60}$ fullerene: 
\begin{equation}
V(r)=\left\{ \begin{array} {ll}
-V_0,  &    {r_i<r<r_o}  \\
0,     &    {r<r_i\ or\ r>r_o}
\end{array}
\right.
\end{equation}
where $V_0=0.3028$ a.u. [20], which is obtained from the experimental electron  
affinity energy for the negative C$^-_{60}$ ion [26] and  
a spherical shell model potential for the C$_{60}$ electrons [14]. 

Equation (1) is a second order differential equation. The general solution is  
a linear combination of a regular solution $u_{kl}(r)$ and an irregular solution 
$v_{kl}(r)$. In this paper the regular solution is obtained from reference [27]. 
The 
irregular solution is evaluated using equation (9.3-23) of reference [28].  
In the region $r<r_i$, $r_i<r<r_0$ and $r>r_0$ the general solutions of equation (1) are
given by:  
\begin{equation}
P(r)=\left\{ \begin{array} {ll}
A* u_{kl}(r)  &    {r<r_i}  \\
B*u_{ql}(r)+C*v_{ql}(r)     &    {r_i<r<r_0} \\
u_{kl}(r)* cos\delta_l -v_{kl}(r)* sin\delta_l  & {r>r_0} 
\end{array}
\right.
\end{equation}
where $q=\sqrt{k^2+K_0^2}$, $K_0^2=2V_0$ and $\delta_l$ is the phase shift due to the 
photoelectron scattering by the potential of the C$_{60}$ fullerene.
$u_{kl}$ is the wave function corresponding to $V(r)=0$, which can be obtained 
by the package of 
Ref. [27]. For the Xe 4$d -\epsilon f$ transition we create three excited states and 
forty continuum wave functions. Each radial wave function was 
represented by 1000 points.
The intrashell correlations can occur among all these excited and ionized states.
$u_{ql}$ is treated similarly; however, $k^2$ is replaced by $q^2$ in Eq. (1).

The coefficients A, B, C, cos$\delta_l$,
and sin$\delta_l$ in equation (3) have been obtained through the use of the  
continuous boundary conditions of the wave functions and their logarithmic derivatives
at $r=r_i$ and $r=r_0$, leading to 

\begin{equation}
B={u_{kl}(r_0) cos\delta_l-v_{kl}(r_0) sin{\delta_l} \over
u_{ql}(r_0)+{D\over F} v_{ql}(r_0)}
\end{equation}
\begin{equation}
A=B*{u_{ql}(r_i)+{D \over F} v_{ql}(r_i) \over u_{kl}(r_i)}
\end{equation}
\begin{equation}
C={D \over F}*B
\end{equation}
\begin{equation}
sin\delta_l={G\over \sqrt{G^2+H^2}}
\end{equation}
\begin{equation}
cos\delta_l={H \over \sqrt{G^2+H^2}}
\end{equation}
where $D$, $F$, $G$ and $H$ are given by 
\begin{equation}
D=u_{kl}(r_i)u_{ql}'(r_i)-u_{kl}'(r_i)u_{ql}(r_i)
\end{equation}
\begin{equation}
F=u_{kl}'(r_i)v_{ql}(r_i)-u_{kl}(r_i)v_{ql}'(r_i) 
\end{equation}
\begin{equation}
G=(u_{ql}(r_0)+{D\over F}v_{ql}(r_0))*u_{kl}'(r_0)-(u_{ql}'(r_0)+{D\over F}v_{ql}'(r_0))*u_{kl}(r_0)
\end{equation}
\begin{equation}
H=(u_{ql}(r_0)+{D\over F}v_{ql}(r_0))*v_{kl}'(r_0)-(u_{ql}'(r_0)+{D\over F}v_{ql}'(r_0))*v_{kl}(r_0)
\end{equation}
The symbol $(')$ means taking the derivative with respect to $r$.
After creating the wave functions, 
the random phase approximation with exchange method [27] was 
used to obtain the photoionization cross sections for the Xe $4d$, $5s$ and $5p$ electrons
in the Xe@C$_{60}$ endohedral molecule.
The cross sections are also evaluated by the approximate formula [11, 12]
\begin{equation}
\sigma_{Xe@C_{60}}=A^2 * \sigma_{Free-atom}.
\end{equation}
where $\sigma_{Free-atom}$ is the photoionization cross section for a free atom.

\section{Results}
Figure 1 shows the  
photoionization cross sections versus photon energy  
for the Xe $4d$ electron of the Xe@C$_{60}$ endohedral molecule.
The solid and dashed curves are respectively, the results 
from the RPAE calculation and equation (13). The calculations included only the $4d -\epsilon f$ channel .
Both calculations are reasonably  
close to each other.
Similar results were found for the Xe $5s$ and $5p$ electrons of the Xe@C$_{60}$
endohedral molecule. 
This indicates that equation (13) is
a good approximation to the RPAE calculation
if only intrashell correlations are included in the calculation. 

The peak at 92.7 eV is partly caused by the reflection effect of 
the C$_{60}$ potential.  
The reflected wave combined with the initial wave reaches a large maximum inside the C$_{60}$.
At photon energy of 92.7 eV the combined radial part of the $\epsilon f$ wave function has a maximum  
which is about 2.2 times larger than that of the same wave function but photoionized from 
the free Xe atom.  Therefore the dipole matrix element reaches a maximum 1.9, at 92.7 eV. 

Using equation (13) implies that we consider the photoionization processes occuring only inside
the C$_{60}$ shell. However, 
the RPAE calculation involves the intrashell correlations 
occuring both inside and outside of the C$_{60}$. Therefore the peak, which shifts 
from 92.7 eV (dashed curve obtained from equation (13)) to 90.9 eV (solid curve, 
RPAE) in Fig. 1, 
demonstrates the importance of the intrashell  
correlations occuring outside the C$_{60}$.  

Figure 2 displays the Xe $4d$ photoionization cross section versus 
photon energy when five channels, Xe $4d-\epsilon f$, $4d-\epsilon p$,
$5s-\epsilon p$, $5p-\epsilon s$ and $5p-\epsilon d$ are included in the RPAE calculation.  
The solid curve and the dotted  curve represent respectively, the Xe $4d$ 
photoionization from the Xe@C$_{60}$ endohedral molecule and the free Xe atom.
The black dots are the experimental data for the free Xe atom [29]. 
The dotted curve has a large and broad maximum, so called the giant resonance. 
The peak of the solid curve in Fig. 1 
has been reduced from 121.5 Mb to 65.4 Mb because of the intershell
correlations among the $4d$, $5s$ and $5p$ sub-shells.  Since the giant 
resonance of the $4d-\epsilon f$ transition is a shape resonance, the resonance amplitude 
is mainly
determined by the shape of the potential and will not be greatly affected by the intershell 
correlations in the photoionization of the free Xe atom. The dashed curves in Fig.1 and Fig.2 
have changed a little, the peak is reduced from 137.7 Mb in Fig. 1 to 117.3 Mb
in Fig. 2.  Therefore by comparing 
the dashed curve, which is the result of equation (13), and the solid curve of Fig. 2
we found that  
equation (13) is no longer a good approximation to the RPAE calculation.  
A method, which incorporates the intershell correlation in the calculation,
such as the RPAE has to be employed to evaluate the 
peak position and the amplitude of the cross section in the multichannel calculation.   

Figure 3 compares the photoionization cross sections versus photon
energy for the Xe $4d$ electron confined in C$_{60}$.
The solid, dashed and dotted curves represent, respectively the results
from this paper, the calculation of $\delta$-like potential, and other spherical
, short range attractive well [18].
The $\delta$-like potential and our results have similar resonance structure
but different peak positions and amplitudes.
Ref. [18] under estimates the resonance effect. 
The calculation using delta-like potential correctly solved the Schr\"{o}dinger equation
using both regular and irregular solutions in the region outside the 
C$_{60}$. However, their $\delta$-like potential used an infinitely thin spherical layer,
which is geometrically not a good approximation to the C$_{60}$ shell to describe
the C$_{60}$ fullerene.  
The photoionization cross sections obtained using this 
model are evaluated by an equation similar to equation (13) of this paper. Therefore,
their cross sections do not include the intershell correlations within the confinement condition.
Ref. [18] improved on the $\delta$-like potential by using a potential, $V_0$ for
$r_0>r>r_i$ and 0 otherwise. However, their solutions of the Schr\"{o}dinger equation did not 
include the irregular solution ($\infty$ at the origin) in the region outside of the C$_{60}$ and
in the C$_{60}$ shell. Therefore their wave function could not reflect the phase shift due to the 
photoelectron scattered by the C$_{60}$ potential well [12]. 

 Our calculation used a model potential $V_0$ for $r_0>r>r_i$, otherwise 0; and the 
regular (0 at the origin) and irregular ($\infty$ at the origin) solutions as well to 
solve the Schr\"{o}dinger equation in both the C$_{60}$ shell and outside the shell. 
After creating the wave function a RPAE calculation was performed to obtain 
the photoionization cross section. Our method surmounted the weaknesses 
of both the $\delta$-like potentical and the other spherical short range attractive 
well [18].   

Figure 4 shows the photoionization cross sections for the 
Xe $5s$ electron. 
Curves in Fig. 4 have the same meaning as in Fig. 2.  
When a $5s$ electron interacts with the $4d$ electron
in the energy range far from the $5s$ threshold, where the cross section is
already small, the $4d$ giant resonance, which has a large and broad
cross section with a peak of 121.5 Mb, at 90.9 eV, causes the $5s$ photoionization 
cross section to 
reach the peak of 6.4 Mb at 91.6 eV. The results of equation (13)  
and of the free Xe atom   
are both much smaller than that of the solid curve. 
The solid and the dashed curves
are not close to each other; particularly the dashed curve does not have a 
large 
peak around 91.6 eV. This again demonstrates the failure of equation (13) if 
there exists strong intershell correlations in the photoionization processes. 
Therefore, similar equations in Refs. [11, 12] and their applications in the $\delta$-like 
potential [13-17] should not be expected to predict the correct resonance character 
in the confinement multichannel situation.

Figure 5 compares the photoionization cross sections versus photon
energy for the $5s$ electron of 
the Xe atom
confined in a C$_{60}$ fullerene. 
The curves have the same meaning as in Fig. 3. 
The dashed curve from a $\delta$-like model does not have the peak around 91.6 eV. 
This is because their formula [14] to calculate the photoionization cross section can not 
include strong coupling between the Xe 4$d-\epsilon f$ and 5$s-\epsilon s, \epsilon d$ transition 
in the confinement condition. Ref. [18] also under estimates the confined resonance effect
as their $4d$ photoionization has a little peak around 90 eV.
By comparing
Fig. 1 of Ref. [14] and Fig. 2 of Ref. [18] we found that a second maximum around
50 eV has been missed in Ref. [18]. This might indicate that more points may be needed 
in that calculation. 
Since the resonance is sharp and
only exists in a very narrow energy region a careful numerical calculation with sufficient 
points has to be perfomed to obtain the correct resonance structure.
Because of this it might not be surprising that several authors may plot 
the different resonance structures with different peak positions and amplitudes 
when their calculations use different
energy points.

Figure 6 shows the photoionization cross sections for the Xe $5p$ electron.
The RPAE calculation involves all five channels. Curves have the same meaning as in Fig. 4.
The dotted curve has a broad maximum in the energy region of the Xe $4d$ 
giant resonance, which is similar to the situation of the $5s$ electron in Fig. 4 and
is caused by the intershell coupling with the 
Xe $4d-\epsilon f$ channel.   
Both the solid and the dashed curves show the confinement resonances.
However, the amplitudes and
the positions of the confinement resonances are quite different in the two calculations, particularly
the dashed curve does not have the characteristic peak of 4.2 Mb around 89.5 eV.

\section{Conclusion}

In conclusion, we have performed RPAE calculations for the
$4d$, $5s$ and $5p$ photoionization of the Xe atom  
confined in a C$_{60}$ fullerene. Our method has surmounted the weaknesses of 
both the $\delta$-like potential [11-17]
and the other spherical short range attractive well [18]. 
The comparison with those of previous model potential calcualtions 
demonstrated significantly stronger correlated confinement resonances for the Xe@C$_{60}$
fullerene.
The comparison also shows that 
equation (13) is a suitable approximation for the photoionization process
if only intrashell correlations are important.
In the multichannel calculation, the RPAE method or other similar methods should be employed
to obtain the correct resonance character for the atom confined in a C$_{60}$ 
fullerene.  

\bigskip
\centerline{\bf ACKNOWLEDGMENTS}
\bigskip

This work was supported by the U.S. DOE, Division of Chemical Sciences,
Geosciences and Biosciences, Office of Basic Energy Sciences, Office of Energy Research.   

\bigskip

\begin{enumerate}
\item Wendin G  and W\"{a}stberg B 1993
Phys. Rev. B {\bf 48} 14764 
\item Decleva P, Alti G De, Fronzoni G, Stener M 1999 
J. Phys. B:At.Mol.Opt.Phys. {\bf 32} 4523 
\item Amusia M Ya, Baltenkov A S, Dolmatov V K, Manson S T, and Msezane A Z 2004 
Phys. Rev. {\bf 70} 023201 
\item Varma H R, Deshmukh P C, Dolmatov V K, and Manson S T 2007
Phys. Rev. A {\bf 76} 012711 
\item Lo S, Korol A V, and Solov'yov A V 2007
J. Phys. B:At.Mol.Opt.Phys. {\bf 40} 3973 
\item Puska M J and Nieminen R M 1993
Phys. Rev. A {\bf 47} 1181; -  1994 Phys. Rev. A {\bf 49} 629 
\item Madjet M E, Chakraborty H S, and Manson S T 2007 
Phys. Rev. Lett. {\bf 99} 243003 
\item Neek-Amal M, Tayebirad G and Asgari R 2007
J. Phys. B:At.Mol.Opt.Phys. {\bf 40} 1509 
\item Lyras A and Bachau H 2005
J. Phys. B:At.Mol.Opt.Phys. {\bf 38} 1119 
\item Katayanagi H, Kafle B P, Kou J, Mori T, Mitsuke K, Takabayahi Y, 
Kuwahara E, and Kubozono Y 2008
J. Quant. Spec. Rad. Trans. {\bf 109} 1590 
\item Baltenkov A S 1999 
Phys. Lett. A {\bf 254} 203 
\item Baltenkov A B 1999 
J. Phys. B:At.Mol.Opt.Phys. {\bf 32} 2745 
\item Amusia M Ya, Baltenkov A S, Chernysheva L V, Felfli Z and Msezane A Z 2005
J. Phys. B:At.Mol.Opt.Phys. {\bf 38} L169 
\item Amusia M Ya, Baltenkov A S, and Becker U 2000 
Phys. Rev. A {\bf 62} 012701 
\item Amusia M Ya and Baltenkov A S 2006 
Phys. Rev. A {\bf 73} 062723
\item Amusia M Ya 2006 Phys. Rev. A {\bf 73} 063206
\item Amusia M Ya, Baltenkov A S, and Chernysheva L V 2007
Phys. Rev. A  {\bf 75} 043201 
\item Dolmatov V K and Manson S T 2008 
J. Phys. B:At.Mol.Opt.Phys. {\bf 41} 165001 
\item Connerade J P, Dolmatov V K, and Manson S T 2000 
J. Phys. B:At.Mol.Opt.Phys. {\bf 33} 2279 
\item Connerade J P, Dolmatov V K, Lakshmi P A, and Manson S T 1999 
J. Phys. B:At.Mol.Opt.Phys. {\bf 32} L239 
\item Dolmatov V K and Manson S T 2006 
Phys. Rev. A {\bf 73} 013201 
\item Murphy T A, Pawlik Th, Weidinger A, Hohne M, Alcale R and Spaeth J M 1996 
Phys. Rev. Lett. {\bf 77} 1075   
\item Komatsu K, Murata M, and Murata Y 2005 
Science {\bf 307} 238 
\item Yannoni C S, Hoinkins M, de Vries M S, Bethune D S, Salem J R,
Crowder M S and Johnson R D 1992 Science {\bf 256} 1191 
\item Xu Y B, Tan M Q, and Becker U 1996 
Phys. Rev. Lett. {\bf 76} 3538 
\item Tosatti E, and Manini N 1994 
Chem. Phys. Lett. {\bf 223} 61 
\item Amusia M Ya and Chernysheva L V 1998 
{\it Computation of Atomic Processes} (IOP, Bristol)
\item Korn G A and Korn T M 1961 
{\it Mathematical Handbook for Scientists and Engineers} (McGraw-Hill, New York)
\item K\"{a}mmerling B, Kossman H and Schmidt V 1989 
J. Phys. B:At.Mol.Opt.Phys. {\bf 22} 841 
\end{enumerate}

\leftline{\bf Figure Captions}
Fig. 1. Comparison of our 
RPAE results (solid curve) with the data evaluated with equation (13)
(dashed curve) for the Xe $4d$ photoionization in the Xe@C$_{60}$ 
endohedral molecule when only the $4d-\epsilon f$ channel is considered 
in the scattering processes. 

Fig. 2. Effects of the intershell correlations in the photoionization of the Xe $4d$ electron.
The solid, dashed and dotted curves represent respectively, 
the $4d$ photoionization cross sections in the Xe@C$_{60}$ endohedral molecule 
calculated by the RPAE method and equation (13) and for the free Xe atom evaluated 
by the RPAE calculation
when the five channels are 
included in the calculation. The black dots are the experimental data [29] for
the $4d$ photoionization of the free Xe atom. 

Fig.3. Comparison of our RPAE results (solid curve) with those of the $\delta$-like
potential (dashed curve) and Ref. [18] (dotted curve) for the 
$4d$ photoionization in the Xe@C$_{60}$ endohedral molecule.

Fig. 4. Same as in Fig. 2 but for the $5s$ photoionization. 

Fig. 5. Same as in Fig. 3 but for the $5s$ photoionization. 

Fig. 6. Same as in Fig. 2 but for the $5p$ photoionization.

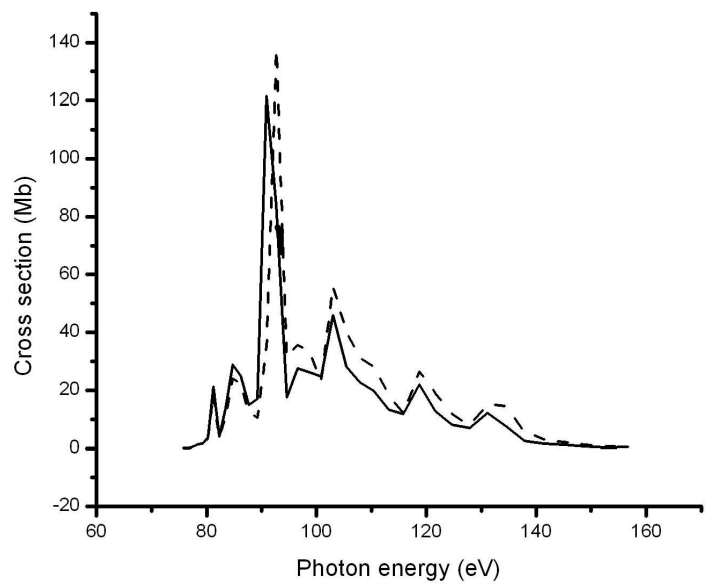
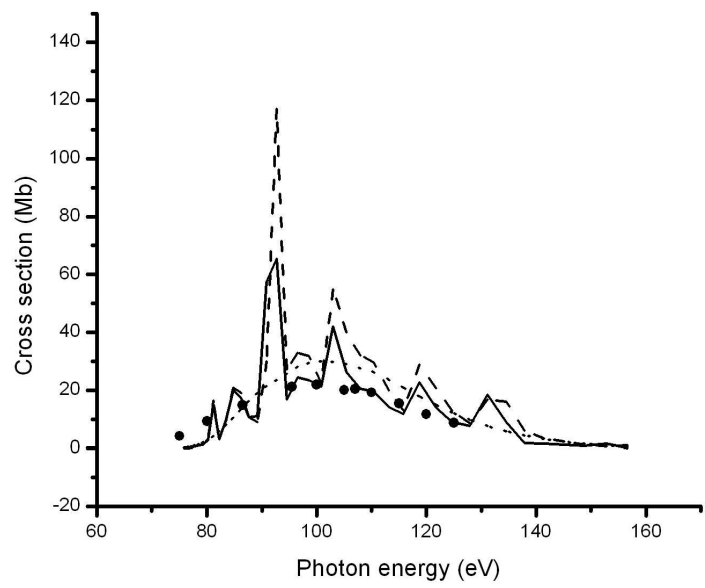
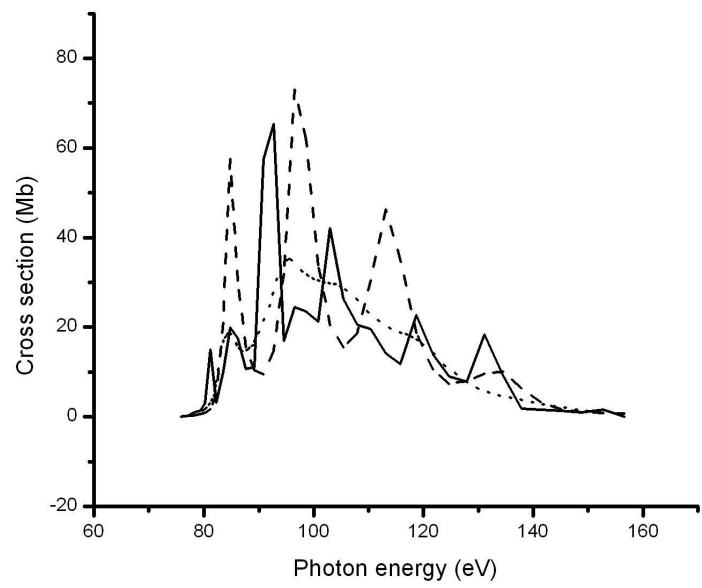
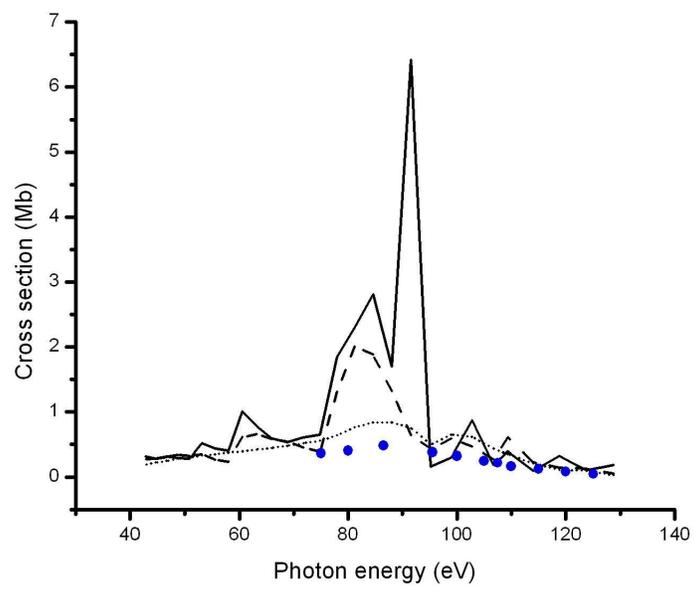
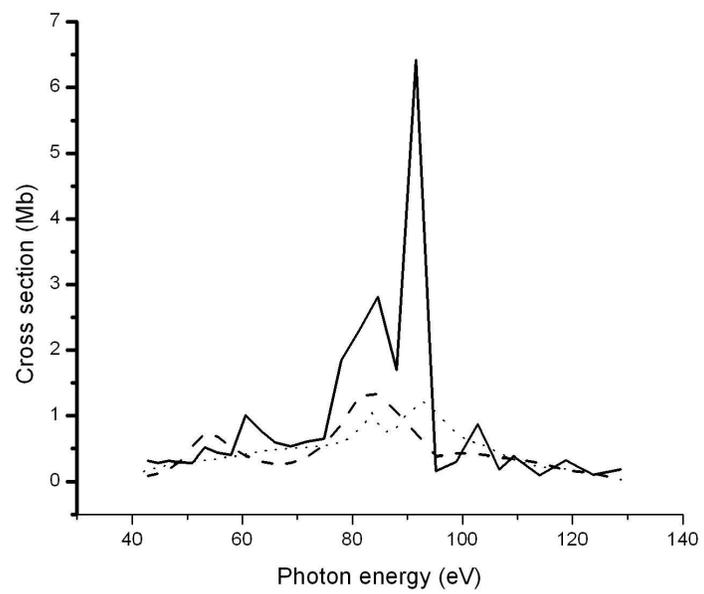
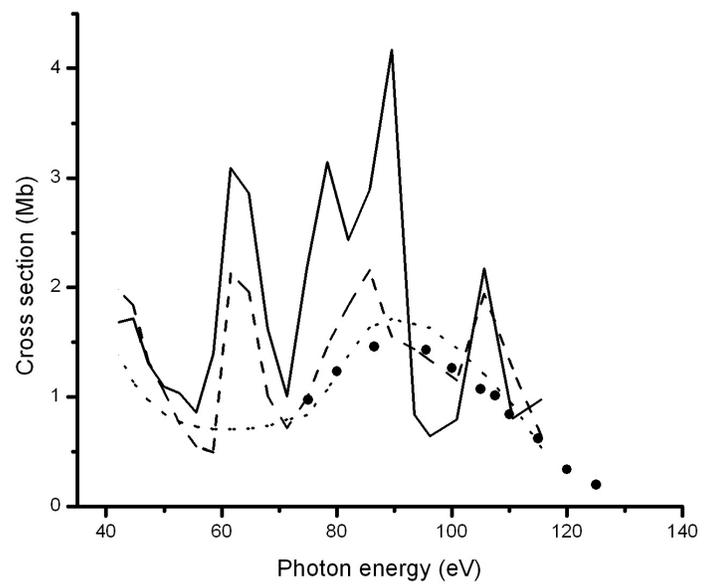
\end{document}